\documentclass[aps,pra,reprint,superscriptaddress]{revtex4-2}

\usepackage[utf8]{inputenc}
\usepackage{amsthm}
\usepackage{amsmath}
\usepackage{amssymb}
\usepackage{physics}
\usepackage{color}
\usepackage{hyperref}
\usepackage{graphicx,xcolor}
\usepackage{caption}
\usepackage{subcaption}

\newcommand{\diff}{\mathrm{d}} 
\newcommand{\mean}[1]{\mathbb{E}\left[#1\right]}
\renewcommand{\tr}{\operatorname{Tr}}

\newcommand{\nket}[1]{|#1\rangle} 
\newcommand{\cavg}[1]{\langle#1\rangle} 
\newcommand{\bvec}[1]{\boldsymbol{#1}} 

\begin{document}
\title{nuHOPS: A quantum trajectory method for highly excited environments in non-Markovian open quantum dynamics}

\author{Kai Müller}
\affiliation{Institute of Theoretical Physics, TUD Dresden University of Technology, 01062, Dresden, Germany}
\author{Walter T. Strunz}
\affiliation{Institute of Theoretical Physics, TUD Dresden University of Technology, 01062, Dresden, Germany}

\begin{abstract}
Systems in contact with an environment provide a ubiquitous challenge in quantum dynamics. Many fascinating phenomena can arise if the coupling is strong, leading to non-Markovian dynamics of the system, or collective, where the environment can become highly excited.
We introduce a 
significant improvement of the
Hierarchy of Pure States (HOPS) approach, which is able to efficiently deal with such highly excited, non-Markovian environments in a nearly unitary way.
As our method relies on quantum trajectories, we can obtain dynamics efficiently, also for large system sizes by i) avoiding the quadratic scaling of a density matrix and ii) exploiting the localization properties of the trajectories with an adaptive basis.
We provide the derivation of the nuHOPS (nearly unitary Hierarchy of Pure States) method, compare it to the original HOPS {and} discuss numerical subtleties based on an illustrative dephasing model. Finally, we show its true power using the Dicke model as the paradigmatic example of many emitters decaying superradiantly inside a cavity. We reach numerically exact solutions for up to 1000 emitters.
We apply our method to study emerging higher order correlations in the emitter system or the cavity mode environment and their scaling with the number of emitters.
\end{abstract}
\maketitle

\section{Introduction}
The efficient numerical simulation of open quantum system dynamics remains a formidable and important challenge of modern quantum physics.
This is especially true for non-Markovian environments \cite{Review_de_Vega} which appear 
frequently
in quantum optics \cite{GardinerNoise}, chemical physics \cite{Citty2024Apr,Roden2012Nov,Potocnik2018Mar}, solid state physics \cite{Weiss2011Nov,Rotter2015Oct,Groblacher2015Jul} and their intersections \cite{Francesco_nonthermal_steady_state}.
In addition, in recent years there has been a growing number of experiments in which the environment can also become highly excited during the dynamics, either due to strong coupling \cite{ultrastrong_coupling_review} or collective effects. 
The paradigmatic example for the latter is Dicke superradiance \cite{Dicke_orig_work}, when the cavity modes are considered as part of an overall non-Markovian environment of the atomic ensemble. Then, the coupling of many emitters to such a shared environment can lead to the emission of an intense, superradiant, burst of radiation and thus highly excite the surrounding environment. 
These collective effects have been observed in a wide variety of physical platforms ranging from ultracold gases \cite{superradiance_BEC, superradiance_lattice_uc_atoms, superradiance_Rydberg} to superconducting qubits \cite{superradiance_circuit_QED}, quantum dots \cite{superradiance_quntum_dots, superradiance_quntum_dots_lattice} and two dimensional materials \cite{superradiance_2D_mat}.\\
While some of these experiments can be described by perturbative or cumulant expansion \cite{Cumulants_Yelin,free_space_SR_cumulant_ok} methods, this is not always the case. 
Especially in the case of strong coupling or when one is interested in quantum properties or non-Gaussian correlations of the system \cite{Stitley_TC, PRL_non_Gauss_driven_cloud} more sophisticated approaches are required. 

Methods that provide a numerically exact solution for non-Markovian open-quantum systems can be roughly divided into two groups: methods like the multilayer multiconfiguration time-dependent Hartree (ML-MCTDH) \cite{ML-MCTDH} or the Davydov-Ansatz \cite{Davydov_review,Davydov_Frank} discretize the environment as a large but finite set of harmonic oscillators and then use variational principles to find a solution for the full quantum state. In these approaches, the environmental degrees of freedom need to be represented explicitly, setting bounds for the applicability due to memory requirements.
By contrast, the second group works on the level of the reduced system state directly and tries to replace the actual environment by some minimal effective environment that is either generated automatically from the influence functional as in the TEMPO and uniTEMPO method \cite{TEMPO, TEMPO_Valentin} or found by a fitting procedure of the bath correlation function as in the Hierarchical equations of motion (HEOM) \cite{HEOM}.
As these methods are based on density matrices they work best for small systems like the spin-boson or few-level models and become expensive when the system itself becomes large, as in the case of many collective emitters. 
The Hierarchy of Pure States (HOPS) \cite{HOPS, HOPS_Richard} approach, is yet a third approach, as it arises directly from a stochastic pure state representation of the exact (full) dynamics and can be seen as a quantum trajectory equivalent of the HEOM equations \cite{PRXQuantum.3.020348}. Clearly, HOPS only scales with the size of the state vector instead of the density operator. 

So far, all numerically exact open system methods mentioned earlier cannot deal satisfactorily with highly excited environments. This is also true
for the minimal effective environments used in HEOM as well as in HOPS: they are unsuitable in dealing with highly excited environments as the convergence of the hierarchy cannot be assured.
Here we rectify this issue by introducing a new co-moving effective environment for the HOPS method, which ensures an optimal expansion into auxiliary states at all times. Remarkably, in contrast to the HOPS, 
these novel extensions can be brought in a form such that the quantum trajectory evolution turns into a nearly unitary Hierarchy of Pure States (nuHOPS).
Our result thus provides an algorithm that is able to efficiently deal with large open systems ($\operatorname{dim} \mathcal{H}_s \approx 1000$) in contact with non-Markovian reservoirs that can also become highly excited. 
Due to the optimal expansion with respect to the hierarchy order, one could also expect an improved efficiency for problems, where HOPS has previously been applied to \cite{asymp_entanglement_Richard,valentinBoettcher,Gera2023May,HOPS_MPS_Stuart}.\\
In this contribution we present a detailed derivation of the nuHOPS method and compare it to the original HOPS \cite{HOPS, HOPS_Richard}, and to master equation approaches, where possible. 
We use our algorithm to find an exact numerical solution for the (Hepp-Lieb) Dicke Model of many emitters inside a single mode cavity for an unprecedented number of emitters up to $N\approx 1000$.
As the HOPS method gives us access to the full state of system and environment, we are able to investigate the occurrence of non-Gaussian correlations between the emitters as well as in the output light.\\
In an accompanying letter we use our method to find genuine quantum effects in a generalized (unbalanced) Dicke Model at large $N$.\\
This article is structured as follows.
We start by deriving our nuHOPS method in Sec.~\ref{sec:method}, where we also compare its numerical efficiency to the original HOPS. In Sec.~\ref{ssec:dephasing} we investigate the method using a toy model first. In this way we can study in detail its numerical implications and subtleties. In particular, we discuss at length a rare stability issue, which manifests itself predominantly for large systems, and show how to cure it. The same instability also occurs in the original HOPS and likely other trajectory methods that rely on non-Hermitian effective Hamiltonians. After describing how the formalism can be extended to different initial states in Sec.~\ref{sec:finite_T}, we turn to the aforementioned example of superradiant decay in the (Hepp-Lieb) Dicke Model in Sec.~\ref{sec:sr_decay}. 

\section{Hierarch of Pure States and its near-unitary reformulation: nuHOPS}\label{sec:method}
We consider a quantum system in contact with an environment, which we model as a collection of infinitely many harmonic oscillators that are coupled linearly to the system. The total Hamiltonian is thus written in the form
\begin{equation}\label{eq:Htot}
    \begin{split}
        H_{tot} &= H_{sys} + H_{int} + H_{bath}\\
            &= H_{sys} + \sum_\lambda \left(g_\lambda^*La_\lambda^\dagger + g_\lambda L^\dagger a_\lambda \right) + \sum_\lambda\omega_\lambda a^\dagger_\lambda a_\lambda,
    \end{split}
\end{equation}
where $a_\lambda$($a_\lambda^\dagger$) are bosonic annihilation (creation) operators for the respective modes, $H_{sys}$ determines the free evolution of the system and the coupling to the modes at frequency $\omega_\lambda$ is described by the system coupling operator $L$ and coupling strength $g_\lambda$.
In the limit of a continuous number of modes the  bath can be described by its spectral density $J(\omega) = \sum_\lambda \abs{g_\lambda}^2 \delta(\omega-\omega_\lambda)$ or equivalently by its bath correlation function $\alpha(\tau) = \int_0^\infty J(\omega)e^{-i\omega \tau} \diff{\omega}$. We first show the derivation for the zero temperature case, where the total initial state is assumed to be a product between an arbitrary initial system state $\ket{\psi_0}$ and the ground state of the bath $\ket{\Psi(0)} = \ket{\psi_0}\otimes\ket{0}$. The generalization to finite temperatures is presented in Sec.~\ref{sec:finite_T}. With this open system model, our goal is to determine the dynamics of the reduced system state $\rho_{sys}(t) = \tr_{bath}\left(\ket{\Psi(t)}\!\bra{\Psi(t)}\right)$.\\
The HOPS method is a numerical implementation of the non-Markovian Quantum State Diffusion (NMQSD) equation \cite{NMQSD}.
For an accessible presentation, we start by sketching the derivation of the NMQSD equation, full details can be found in \cite{NMQSD}. By taking the trace in the basis of (unnormalized) Bargmann coherent states $\ket{|z_\lambda} = \exp(z_\lambda a_\lambda^\dagger) \ket{0}$ we can express the reduced system state as
\begin{equation}\label{eq:linear_mean}
    \begin{split}
        \rho_{sys}(t) =& \int \left(\prod_\lambda \dd^2{z_\lambda} \,e^{-\abs{z_\lambda}^2}\right)\bra{z_\lambda |}\ket{\Psi(t)}\!\bra{\Psi(t)}\ket{|z_\lambda}\\
            =& \mean{\bra{\vec{z}|}\ket{\Psi(t)}\!\bra{\Psi(t)}\ket{|\vec{z}}}\\
            \equiv& \mean{\ketbra{\psi_t(z^*)}},
    \end{split}
\end{equation}
where in the second step we have interpreted the integral as a classical expectation value with respect to Gaussian random variables $z_\lambda$ and $\ket{\vec{z}} = \bigotimes_\lambda \ket{\vec{z}_\lambda}$. To obtain the time evolution of $\ket{\psi_t(z^*)}=\bra{\vec{z}|}\ket{\Psi(t)}$ we switch to an interaction picture with respect to the free bath Hamiltonian 
\begin{equation}\label{eq:HtotI}
    \begin{split}
        H_{tot}^I &= H_{sys} + iB^\dagger(t)L -iB(t)L^\dagger,\\
        B(t) &= i\sum_\lambda g_\lambda e^{-i\omega_\lambda t}a_\lambda.
    \end{split}
\end{equation}
By using the eigenstate property of the (Bargmann) coherent states $a_\lambda\ket{|z_\lambda} = z_\lambda\ket{|z_\lambda}$, we find for the time evolution
\begin{equation}\label{eq:linear_nmqsd}
    \begin{split}
        \partial_t \ket{\psi_t(z^*)} =& -i\bra{\vec{z}|}H_{tot}^I\ket{\Psi},\\
        =& -iH_{sys}\ket{\psi_t(z^*)} + z_t^*L\ket{\psi_t(z^*)} \\
        &- L^\dagger\bra{\vec{z}|}B(t)\ket{\Psi(t)},
    \end{split}
\end{equation}
where 
\begin{equation}\label{eq:linear_nmqsd_noise}
       z_t^* = -i\sum_\lambda g_\lambda^* e^{i\omega_\lambda t} z_\lambda^*.
\end{equation}
Upon considering the $z_\lambda$ as Gaussian random variables, this $z_t^*$ turns into a Gaussian stochastic process
with mean zero and the bath correlation $\mean{z_t z_s^*} = \sum_\lambda \abs{g_\lambda}^2 e^{-i\omega_\lambda (t-s)} = \alpha(t-s)$.
In NMQSD, the non-trivial term $\bra{\vec{z}|}B(t)\ket{\Psi(t)}$ in \eqref{eq:linear_nmqsd} is replaced by a memory integral over a functional derivative with respect to the stochastic process,
\begin{equation}\label{eq:linear_nmqsd_functional_derivative}
  \bra{\vec{z}|}B(t)\ket{\Psi(t)} = 
  \int_0^t\dd{s}\alpha (t-s) \frac{\delta}{\delta{z}_s^*}\ket{\psi_t({z}^*)},
\end{equation}
from which the HOPS can be derived (see Appendix \ref{app:derivation}). Here and in the following, however, we stick to the expression on the left hand side.

Note that the trajectories $\ket{\psi_t({z}^*)}$ are not normalized, such that trajectories with a small norm will only marginally contribute to the ensemble average in \eqref{eq:linear_mean}. 
To improve the sampling efficiency one can perform a Girsanov transformation as shown in Ref.~\cite{NMQSD}, after which the system state is obtained as an average over normalized trajectories $\rho_{sys}(t) = \mean{\ket{\psi_t(\widetilde{z}^*)}\!\bra{\psi_t(\widetilde{z}^*)}/{\braket{\psi_t(\widetilde{z}^*)}}}$. 
With this transformation one obtains the non-linear NMQSD equation
\begin{align}\label{eq:nonlinear_nmqsd}
        \partial_t \ket{\psi_t(\widetilde{z}^*)} =& -iH_{sys}\ket{\psi_t(\widetilde{z}^*)} + \widetilde{z}_t^*L\ket{\psi_t(\widetilde{z}^*)} \\
        &- (L^\dagger -\cavg{L^\dagger}_t)\langle\vec{\widetilde{z}}||B(t)\ket{\Psi(t)},\notag
\end{align}
where
\begin{align}
        \cavg{L^\dagger}_t = \bra{\psi_t(\widetilde{z}^*)}L^\dagger\ket{\psi_t(\widetilde{z}^*)}/{\braket{\psi_t(\widetilde{z}^*)}}.
\end{align}
Here,
\begin{align}
        \label{eq:def_stocproc}
        \widetilde{z}_t^*  =& z_t^* + \int_0^t \alpha^*(t-s)\cavg{L^\dagger}_s\dd{s}
\end{align}
is the Girsanov-transformed process, which is a sum of the original Gaussian process with the bath correlations from \eqref{eq:linear_nmqsd_noise} and a shift that will play a prominent role in nuHOPS soon.

In the following we will derive a new version of the HOPS \cite{HOPS, HOPS_Richard} that allows to determine the trajectories $\ket{\psi_t(\widetilde{z}^*)}$ more efficiently.
For simplicity we first only consider the case of a single exponentially decaying correlation function $\alpha(\tau) = Ge^{-w\tau}$ for $\tau > 0$. 
HOPS can be applied to any bath correlation function by approximating the true $\alpha(\tau)$ by a sum of exponentials $\alpha (\tau) \approx \sum_i G_ie^{-w\tau}$, where $G,w\in \mathbb{C}$ are \textit{arbitrary complex} fit parameters (restricted merely by the condition that $\alpha(\tau)$ has to be a proper bath correlation function).
The generalization of the derivation shown here to this more general case is somewhat unwieldy, but straight forward and presented in Appendix \ref{app:derivation}. \\
From Eq.~\eqref{eq:nonlinear_nmqsd} we proceed along the lines of the original HOPS derivation \cite{HOPS} to express the term $\bra{z_\lambda |}B(t)\ket{\Psi(t)}$ (which gives rise to the functional derivative \eqref{eq:linear_nmqsd_functional_derivative} in NMQSD) via auxiliary states. 

However, in contrast to the original HOPS, we now allow for an (at first arbitrary) shift $\mu(t)$ in the definition of the auxiliary states and set
\begin{equation}\label{eq:defAuxStates}
    \nket{\psi_t^{(k)}}  = \frac{(-i)^k}{\sqrt{k!G^k}}\bra{\vec{z} |}(B(t) - \mu(t))^k\ket{\Psi(t)}.
\end{equation}
The time-dependent function $\mu(t)$ will later be chosen in a way that optimizes numerical efficiency and stability. The original HOPS corresponds to the choice $\mu(t)=0$.
Note that our desired trajectories $\ket{\psi_t(z^*)}$ are included in the above definition as the zeroth order states $|\psi_t^{(0)}\rangle$ (for better readability, we suppress their dependence on the noise $z^*_t$ in the following).\\
A crucial property of an environment with an exponentially decaying bath correlation function $\alpha(\tau) = Ge^{-w\tau}$ at zero temperature is that $\bra{\vec{z}|}(\partial_tB(t))\ket{\Psi(t)} = -w \bra{\vec{z}|}B(t)\ket{\Psi(t)}$ \cite{HOPS}, see Appendix \ref{app:derivation}. With this property and with the auxiliary states defined as above in \eqref{eq:defAuxStates} we find a new hierarchy of pure states
\begin{equation}\label{eq:HOPS_hierarchy}
    \begin{split}
        \partial_t\nket{\psi_t^{(k)}} =& \left(-i H_{sys} + \widetilde{z}_t^*L-\mu(t)L^\dagger -k(i\omega_c+\kappa)\right)\nket{\psi_t^{(k)}}\\
        &-i\sqrt{Gk} \left(L - \left(\frac{\dot{\mu} + (i\omega_c+\kappa)\mu}{G}\right)\right) \nket{\psi_t^{(k-1)}}\\
        &-i\sqrt{G(k+1)}\left(L^\dagger - \cavg{L^\dagger}_0\right)\nket{\psi_t^{(k+1)}}.
    \end{split}
\end{equation}
We emphasise that the expectation value $\cavg{L}_0$ is always taken with respect to the zeroth order hierarchy state $\nket{\psi^{(0)}_t}$. It is convenient \cite{HOPS_MPS_Alex,HOPS_MPS_Stuart} to rewrite the hierarchy with the help of an auxiliary oscillator degree of freedom by introducing the single HOPS state $\ket{\Phi}$ through the Fock states $\ket{k}$ of the auxiliary oscillator by setting
\begin{equation}\label{eq:def_HOPS_state}
    \ket{\Phi(t)} = \sum_k \nket{\psi_t^{(k)}}\ket{k}.
\end{equation}
With the raising (lowering) operators of this auxiliary oscillator $b^\dagger$($b$) the hierarchy can be expressed as
\begin{equation}\label{eq:nuHOPS_raw}
    \begin{split}
        \partial_t \ket{\Phi(t)} &= \left(-i H_{sys} + \widetilde{z}_t^*L -\mu(t)L^\dagger -(i\omega_c+\kappa)b^\dagger b\right)\ket{\Phi(t)}\\
        &-i\sqrt{G} \left(L - \left(\frac{\dot{\mu} + (i\omega_c+\kappa)\mu}{G}\right)\right) b^\dagger\ket{\Phi(t)}\\
        &-i\sqrt{G}\left(L^\dagger - \cavg{L^\dagger}_0\right)b\ket{\Phi(t)}.
    \end{split}
\end{equation}
We can see that in this picture, the hierarchy Eq.~\eqref{eq:HOPS_hierarchy} corresponds to a
Fock state expansion with respect to the auxiliary oscillator. Thus, the hierarchy can only be
numerically efficient if the state of the auxiliary oscillator is 
located close to $\ket{0}$ or, in other words, its Q-function 
is located close to the origin of its phase space. In Appendix \ref{app:optimal_shift} we show, that by choosing a cleverly designed
shift $\mu(t)$, one can always ensure that $\cavg{b} = 0$, i.e. the 
Q-function can indeed be anchored around zero. In practice, calculating this optimal $\mu(t)$ is rather involved. As we will see next, it turns out to be both, more elegant and more efficient, to obtain $\mu(t)$ from a mean-field approximation of the auxiliary oscillator dynamics. This $\mu_{mf}(t)$ is identical to the optimal $\mu(t)$ if the HOPS state \eqref{eq:def_HOPS_state} has the form $\ket{\Phi(t)} = \ket{\psi_t}\otimes\ket{\beta}$, where $\ket{\beta}$ is a coherent state. With this choice one obtains the mean-field optimal shift
equation
\begin{equation}\label{eq:mu_mf_dgl}
    \partial_t\mu_{mf}(t) = -w\mu_{mf}(t) + {G}\cavg{L}_0,
\end{equation}
which is solved along with the HOPS equation. Note that for our exponential bath correlation function its formal solution is just
\begin{equation}\label{eq:mu_mf}
    \mu_{mf}(t) = \int_0^t\alpha(t-s) \cavg{L}_0(s)\dd{s},
\end{equation}
which is equal to the complex conjugate of the stochastic process shift in Eq.~\eqref{eq:def_stocproc}. Thus, both terms can be combined to give a Hamiltonian-like contribution to the right-hand-side of Eq.~\eqref{eq:nuHOPS_raw}: $\mu_{mf}^*(t)L-\mu_{mf}(t)L^\dagger$. In addition, also the terms linear in $b$ and $b^\dagger$ turn out to contribute in a Hamiltonian manner which means that the evolution equation \eqref{eq:nuHOPS_raw} describes nearly unitary dynamics of $|\Phi(t)\rangle$.

Thus, with $\mu(t) = \mu_{mf}(t)$ we obtain the main result of this work, the nearly unitary hierarchy of pure states (nuHOPS), expressed in terms of an
auxiliary oscillator:
\begin{equation}\label{eq:shiftedHOPS}
    \begin{split}
        \partial_t \ket{\Phi(t)} =& -i \Bigg[H_{sys} + \omega_c b^\dagger b +i (\mu_{mf}^*(t)L-\mu_{mf}(t)L^\dagger)\\
 & + \sqrt{G} \Big(\left(L - \cavg{L}_0\right)b^\dagger +  \left(L^\dagger - \cavg{L^\dagger}_0\right)b\Big)\Bigg] \ket{\Phi(t)} \\
        &    + \left({z}_t^*L -\kappa b^\dagger b\right) \ket{\Phi(t)}.
    \end{split}
\end{equation}
We stress that in Eq.~\eqref{eq:shiftedHOPS} it is the last line only that is not of Hamiltonian form, the overall dynamics is near-unitary. Moreover, by construction, its zero-component $\psi_t(\widetilde{z}^*) = \braket{0}{\Phi(t)}$ offers an exact stochastic unravelling of the reduced density operator $\rho_{sys}(t) = \mean{\ket{\psi_t(\widetilde{z}^*)}\!\bra{\psi_t(\widetilde{z}^*)}/\braket{\psi_t(\widetilde{z}^*)}}$ of the full model of Eq.~\eqref{eq:Htot}, as in \eqref{eq:linear_mean}. This holds true here for a bath correlation function that is a single exponential, the genereal case is dealt with in App. \ref{app:derivation}.\\

Eq.~\eqref{eq:shiftedHOPS} is the central result of this work: a near-unitary hierarchy of pure states (nuHOPS) that drastically increases the computational efficiency for highly excited environments. We will demonstrate its power to keep the relevant hierarchy depth at a minimum for a simple dephasing example next, and later for a highly demanding application, the Dicke model. Moreover, there is an important numerical subtlety to consider, which can lead to instabilities in the numerical implementation if it is not properly dealt with. These instabilities occur only for large systems and can also appear in the original HOPS method. We will study in detail these fine intricacies as well as the solution to circumvent them in the next Section using the simple dephasing example.

\section{Illustrative example: numerical efficiency and stability}\label{ssec:dephasing}
\begin{figure}
    \begin{minipage}{0.45\textwidth}
        \centering
             \includegraphics[width=\textwidth]{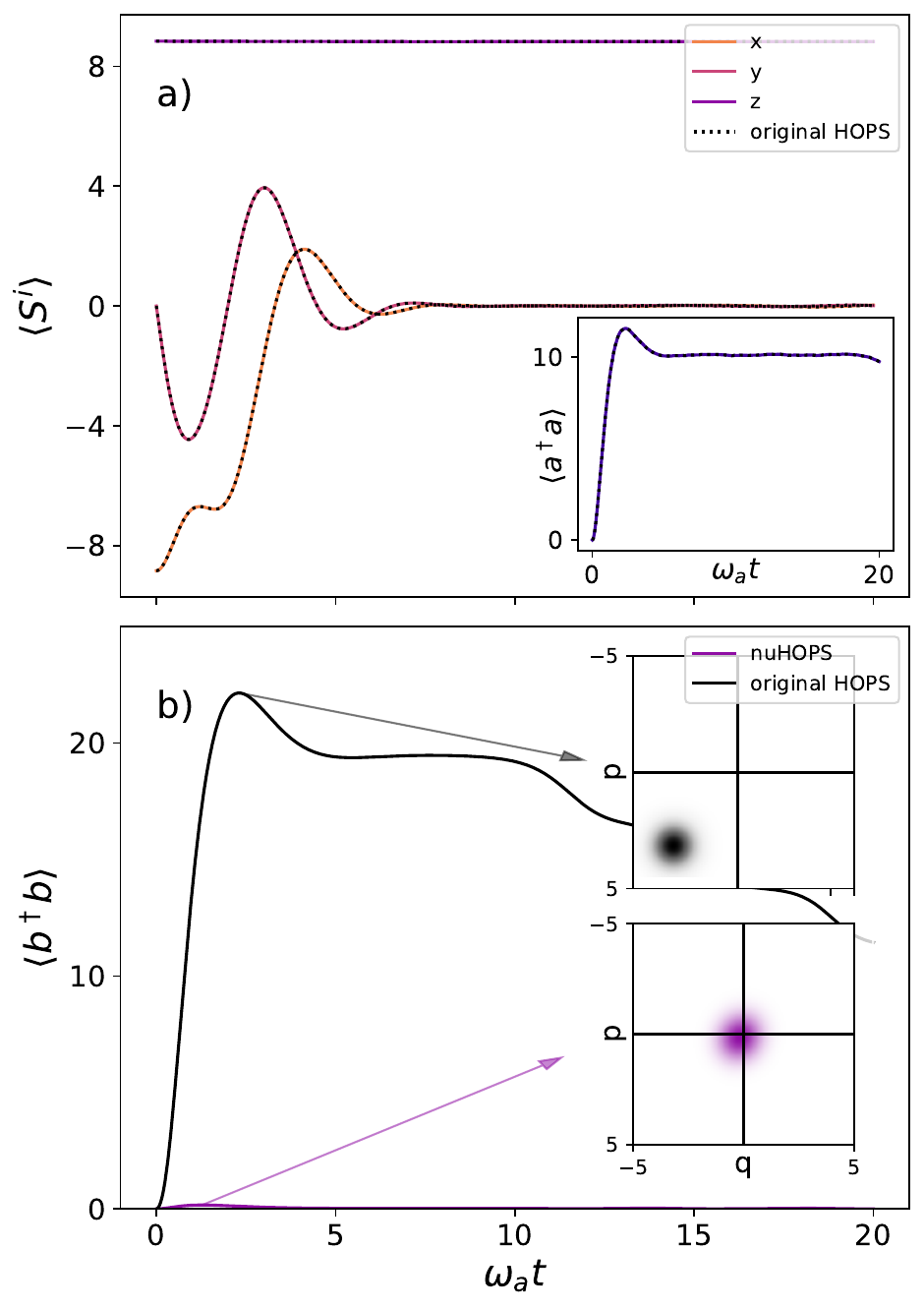}
    \caption{Dynamics of the dephasing model \eqref{eq:MasterDephasing} for $\omega_a = \omega_c = \kappa = 1$, $g=0.5$ and $N=25$. Plot a) shows a perfect agreement in the physical expectation values between the original and the nuHOPS. Plot b) shows how far fewer auxiliary states are required in the nuHOPS, as evidenced by the significantly lower occupation of the auxiliary oscillator. Insets show the Q function of the auxiliary oscillator at the respective times of maximal occupation.}
    \label{fig:dephasing_dynamics}
    \end{minipage}
\end{figure}
After deriving the nuHOPS Eq.~\eqref{eq:shiftedHOPS} we will start by considering a simple dephasing model to gain a better understanding of how the equation works and how it compares to the original HOPS \cite{HOPS, HOPS_Richard}. 
As the dephasing model can be treated analytically to a large extent it will point us to a subtle source of instabilities, which can result in abrupt, seemingly non-analytic jumps of certain trajectories. Although their occurrences are rare, their erratic, singular behavior may spoil the overall ensemble mean and thus they have to be dealt with properly.
The model we consider is that of a spin (dimension $s=N/2$) coupled to a single damped cavity mode, such that the evolution of the total state of system and cavity mode $\rho_{tot}$ is described by the Master equation
\begin{equation}\label{eq:MasterDephasing}
    \begin{split}
        \dot{\rho}_{tot} =& -i[H_{deph},\rho] + \kappa\left(2a\rho a^\dagger - \{a^\dagger a, \rho\}\right),\\
        H_{deph} =& \omega_a S^z + \omega_c a^\dagger a + \frac{g}{\sqrt{N}}S^z\left(a+a^\dagger\right).
    \end{split}
\end{equation}
Here, $S^i$ are spin operators with $[S^i, S^j] = i\epsilon_{ijk}S^k$ and $a$ ($a^\dagger$) is the bosonic annihilation (creation) operator of the cavity mode with $[a, a^\dagger] = 1$. 
In the following we are interested in the reduced evolution of the spin $\rho = \tr_{cavity}\left(\rho_{tot}\right)$. 
As the correlation function of the cavity mode is the single exponential $\alpha (t-s) = \cavg{a(t)a^\dagger (s)} = g^2\exp(-i\omega_c (t-s) - \kappa\abs{t-s})$ we can obtain the reduced state from Eq.~\eqref{eq:shiftedHOPS}, with $H_{sys}=\omega_a S^z$, the coupling operator $L=S^z$ and $G=g^2$.
Thus, the nuHOPS equation can be written as
\begin{equation}\label{eq:shiftedHOPSdeph}
    \begin{split}
        \partial_t \ket{\Phi(t)} =& \Big[-i \omega_a S^z + \left({z}_t^* +\mu_{mf}^*(t)-\mu_{mf}(t)\right)S^z \\
        &-(i\omega_c+\kappa)b^\dagger b\\
        &-ig (\left(S^z - \cavg{S^z}_0\right)\left(b^\dagger +  b\right)\Big]\ket{\Phi(t)}.
    \end{split}
\end{equation}
The dynamics of this model is shown in Figure~\ref{fig:dephasing_dynamics} a) for $s=25/2$ and a spin-coherent initial state $\ket{\psi_0} = \ket{\theta=-\pi/4, \phi=0}$, where $\ket{\theta, \phi} = \exp\left((\theta e^{i\phi} S^- -\theta e^{-i\phi}S^+)/2\right)\ket{j,j}$. 
As expected from a dephasing model, the $z$ component of the spin expectation value stays constant, while the $x$ and $y$ components relax to zero as the state becomes diagonal in the $S^z$ eigenbasis. 
We compute the dynamics both with Eq.~\eqref{eq:shiftedHOPSdeph} as well as with the original HOPS and observe perfect agreement of the reduced system state as well as the cavity photon number expectation values (inset). 
However, while the dynamics of physical quantities are identical, the dynamics of the auxiliary oscillator of the hierarchy differ for the two algorithms. 
As shown in Fig.~\ref{fig:dephasing_dynamics} b) for a single representative trajectory, the expected occupation of the auxiliary oscillator $\cavg{b^\dagger b}$ is close to zero for the nuHOPS, and much smaller than for the original HOPS at all times. Thus, a significantly smaller hierarchy depth is required to capture the same physical dynamics of the open spin and cavity. The insets of Fig.~\ref{fig:dephasing_dynamics} b) show the Husimi-Q function of the auxiliary oscillator for both hierarchies at their maximal occupations.\\
The small occupation of the auxiliary oscillator mode in nuHOPS and the strongly localized Q-function in Fig.~\ref{fig:dephasing_dynamics} suggest that a very small number of hierarchy Fock states is sufficient to accurately represent the state. While this picture holds almost always, very rarely we do find a subtle and surprising instability when solving these equations numerically, which we explore next.

\begin{figure*}
    \begin{minipage}{\textwidth}
        \centering
             \includegraphics[width=\textwidth]{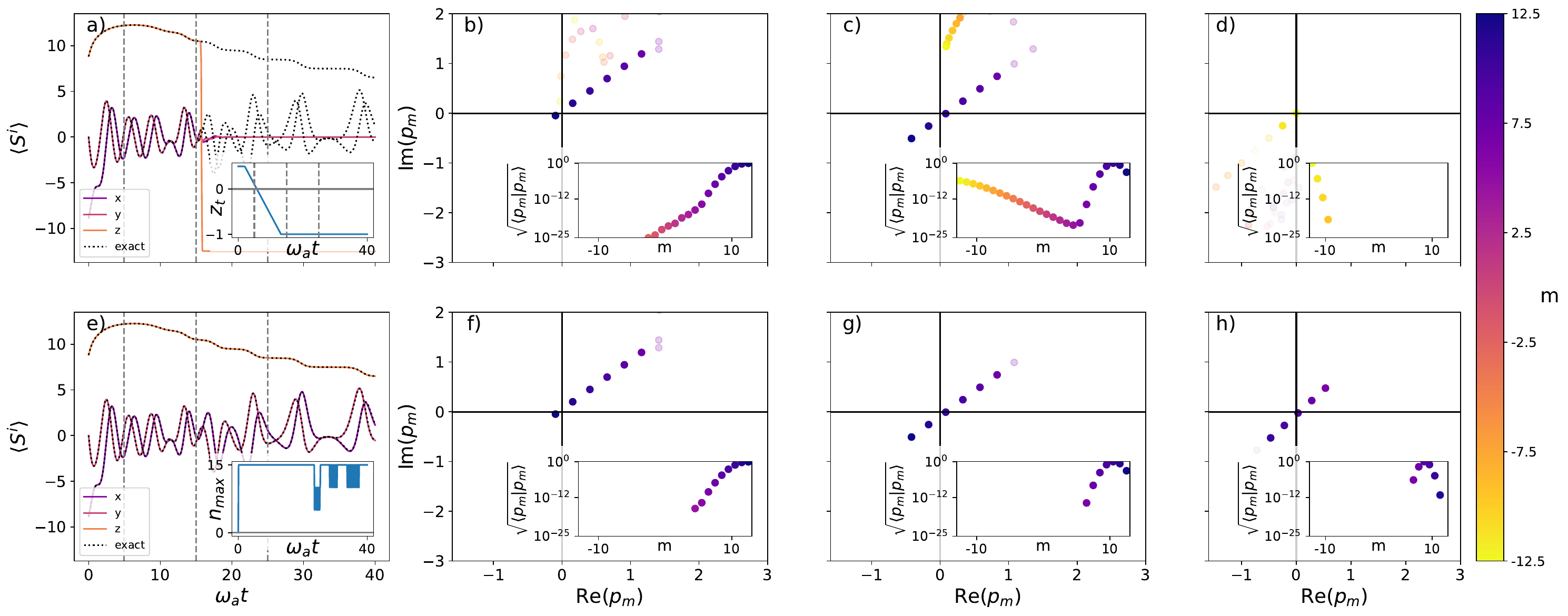}
    \caption{Example of the instabilities and how to prevent them. Plot a) shows the dynamics resulting from a naive implementation of Eq.~\eqref{eq:shiftedHOPSdeph} for the (artificial) noise shown in the inset. The instability manifests itself in a seemingly abrupt jump of $\cavg{S^z}$. Plots b)-d) show a detailed representation of the state at different evolution times, marked by the grey dashed lines in plot a). Expanding the full state according to Eq.~\eqref{eq:defpm} in spin-$S_z$ eigenstates $\ket{s,m}$, the plots show the value of $\bra{p_m}b\ket{p_m}/\braket{p_m}$ in the complex plane, where the value of $m$ is encoded in the color. Insets show the norm of the states $\ket{p_m}$ and states with $\braket{p_m} \leq 10^{-20}$ are shown transparent in the complex plane. Plots e) -h) show the same trajectory calculated with an adaptive basis, which prevents the instabilities (see text).}
    \label{fig:dephasing_adaptivity}
    \end{minipage}
\end{figure*}

Naively solving Eq.~\eqref{eq:shiftedHOPSdeph} and restricting the basis of the auxiliary oscillator to 15 Fock states leads to rare, seemingly non-analytic jumps of some of the trajectories, as shown in Fig.~\ref{fig:dephasing_adaptivity} a). There the expectation values of a single trajectory corresponding to the artificially designed  "noise" $z_t^*$ shown in the inset are plotted (solid lines). These show a numerical instability emerging from a persistent (here negative) real part of the noise $z_t^*$, leading to an abrupt jump around $\omega_a t\approx 15$.  The results are compared to an exact solution obtained from the original HOPS implementation, which required \textit{80} Fock states to converge (black dotted lines). 
In the following we investigate why these instabilities occur and how they can be prevented, such that with these two hierarchy depths (15 vs. 80) nuHOPS and HOPS ultimately give identical results, as shown in Fig.~\ref{fig:dephasing_adaptivity} e).\\
We first write the full (nu-)HOPS state in the $S^z$ eigenbasis $\{\ket{s,m}\}$, with $S^z\ket{s,m} = m\ket{s,m}$ and $\vec{S}^2 \ket{s,m} = s(s+1)\ket{s,m}$
\begin{equation}\label{eq:defpm}
\begin{split}
    \ket{\Phi(t)} =& \sum_m \ket{s,m}\ket{p_m(t)}. 
\end{split}
\end{equation}
The (unnormalized) $\ket{p_m(t)}=\braket{s,m}{\Phi(t)}$ are states in the Hilbert space of the auxiliary oscillator and evolve according to
\begin{equation}\label{eq:evolutionpm}
\begin{split}
    \partial_t \ket{p_m(t)} =& \Big[Z(t) - (i\omega_c + \kappa)b^\dagger b \\
    &-ig(m-\cavg{S^z}_0)(b+b^\dagger)\Big]\ket{p_m(t)},\\
    Z(t) =& (-i\omega_a + z_t^* + \mu^*(t) - \mu(t))*m.
\end{split}
\end{equation}
Since all $\ket{p_m}$ start in the ground state Eq.~\eqref{eq:evolutionpm} can be solved by an (unnormalized) coherent state Ansatz, where the coherent state labels $p_m(t)$ and norms satisfy the evolution equations:
\begin{equation}\label{eq:norms_labels_deph}
    \begin{split}
        \dot{p}_m(t) =& -wp_m(t) -ig(m-\cavg{S^z}_0),\\
         \partial_t \braket{p_m} =&  \left(2\operatorname{Re}(z_t)m - \kappa \frac{\bra{p_m}b^\dagger b\ket{p_m}}{\braket{p_m}}\right)\braket{p_m} + \mathcal{N}(t),
    \end{split}
\end{equation}
where $\mathcal{N}(t)$ arises from an overall normalization of $\ket{\Phi(t)}$ during the numerical evolution and is irrelevant for the following discussion.
We can see that there is an $m$-dependent change of the norms, connected to the real part of the noise $z_t^*$, which is a consequence of the effective Hamiltonian \eqref{eq:shiftedHOPSdeph} being non-Hermitian.  
It depends on a balance between the noise and the occupation of the auxiliary oscillator, conditioned on $m$. One can imagine that a truncated Fock space, which sets an upper bound on the occupation, may disturb this balance, which brings us to the reason of the instabilities.\\
Consider a situation, where $\cavg{S^z}_0$ has been constant for some time. Then the coherent state labels reach a steady state
\begin{equation}\label{eq:pm_steady}
    p_m^s = -ig/w(m-\cavg{S^z}_0),
\end{equation}
where the coherent state labels for each $m$ are evenly spaced out along a line in the complex plane, and are close to zero if $m\approx\cavg{S^z}_0$. 
Thus most of the weight is located close to zero, whereas states corresponding to $S^z$ eigenvalues far from the expectation value, which will typically only have negligible weight, are pushed out, far away from the origin. 
The typical reasoning would now be that, since their weight is negligible, the Fock space can be truncated close to the origin, which would still capture the relevant part of the state. 
However, from Eq.~\eqref{eq:norms_labels_deph} we see that restricting the Fock space would set an upper bound on ${\bra{p_m}b^\dagger b\ket{p_m}}/{\braket{p_m}}$ and the norm of these small, but far out states may no longer be sufficiently damped.
Especially for large systems, where $m\gg 1$ this can lead to a sudden exponential growth of some previously irrelevant states $m$ resulting in the jumps discussed before.\\
An example of this process is shown in Fig.~\ref{fig:dephasing_adaptivity} b) -d). In these plots the expectation values $\cavg{b}$ (corresponding to the coherent state labels $p_m$) for the states $\ket{p_m}$ are shown in the complex plane at different times, where $m$ is encoded in the color. 
States with a norm $\braket{p_m} < 10^{-20}$ are transparent and the insets show the norms for each $m$.
At $\omega_a t = 5$ we have $\cavg{S^z}_0 \approx 12.5$ and $z_t = 1/2$. 
Fig.~\ref{fig:dephasing_adaptivity} b) shows how the coherent state labels are indeed located along a line in the complex plane, but due to the truncated Fock space the $\ket{p_m}$ can no longer be represented correctly for $m<7$. 
The Fock space truncation is nevertheless an excellent approximation as $\braket{p_m} \lessapprox 10^{-20}$ for $m<7$. 
However up to $\omega_a t = 14.16$ (shown in Fig.~\ref{fig:dephasing_adaptivity} b) $z_t$ decreases and eventually becomes negative. 
While for $\operatorname{Re}(z_t) > 0$ states with small $m$ were suppressed by the first term in Eq.~\eqref{eq:norms_labels_deph} they are now enhanced. 
In an exact solution this enhancement would be balanced by their large Fock space occupation $\bra{p_m}b^\dagger b\ket{p_m}/\braket{p_m}$. Yet since this is no longer accurately represented in the truncated Fock space, they erroneously grow in size.
This mechanism results in the norms shown in the inset of Fig.~\ref{fig:dephasing_adaptivity} c). There, one can clearly observe the exponential growth for small $m$, resulting in the sharp jump of the expectation value in Fig.~\ref{fig:dephasing_adaptivity} a). 
This also explains why the issue appears predominantly for large system sizes, where already a small negative value of $\operatorname{Re}(z_t)$ in this situation would be enough to tremendously enhance a part of the state, that is, however, entirely irrelevant physically .
After the erroneous jump, the state remains localized at $\cavg{S^z}_0 \approx 12.5$ as shown in Fig.~\ref{fig:dephasing_adaptivity} d).\\
Having identified the cause of the unphysical instabilities we now turn to the solution how to avoid them.
As the effective Hamiltonian for the trajectories is non-Hermitian, simply truncating auxiliary Fock space alone is not sufficient. We have seen that instabilities only occur for $\ket{p_m}$ which have a negligible norm. 
If we thus truncate the spin system Hilbert space as well, we can simply set those unimportant states to zero and prevent their norms from growing exponentially. 
As we cannot exclude any part of the system Hilbert space a priori, we let the computer do the work and perform this truncation in an adaptive manner. Similar adaptive strategies for quantum trajectories have been used to increase numerical performance in \cite{Schack1995Sep,Gao2019Jun,Varvelo2021Jul}.
By tracking the norm of $\ket{p_m}$ for the largest and smallest $m$ we increase (decrease) the system basis if an upper (lower) threshold $\epsilon_{up}$ ($\epsilon_{low}$) is reached.
We employ the same adaptive scheme for the Fock state basis of the auxiliary oscillator to ensure that enough Fock states are available to accurately represent each $\ket{p_m}$.
The result can be seen in the lower panels Fig.~\ref{fig:dephasing_adaptivity} e) - h). The inset of Fig.~\ref{fig:dephasing_adaptivity} e) shows the number of Fock states used over time. 
The plots show how with the reduced system basis all $\ket{p_m}$ can be represented accurately and thus no instabilities occur. 
Especially the comparison of plot c) to plot g) shows the adaptive basis alleviates us from the need to accurately represent states which are unimportant for the dynamics. 
As the $\cavg{S^z}_0$ expectation value slowly decreases new states are adaptively added to the basis.

\section{Extension to finite temperature and other initial scenarios}\label{sec:finite_T}
Up to now we have considered environments which are initially in the vacuum (zero temperature) state.
In this section we show how the nuHOPS method can be extended to more general initial states.
First, we briefly show how the known stochastic potential method for finite temperature \cite{HOPS_Richard} can also be applied to the nuHOPS. We then describe how the nuHOPS method opens new opportunities to describe initial states, where only a part of the environment has a finite temperature.\\
We start from the full Hamiltonian in the interaction picture Eq.~\eqref{eq:HtotI}, but consider an initial condition of the form $\rho(0) = \ketbra{\psi_0}\otimes\rho_{\beta}$, where $\rho_{\beta}$ is the Gibbs state of the bath at inverse temperature $\beta = 1/(k_BT)$. With the help of the P-representation $\rho_\beta$ may be written as a mixture of coherent states
\begin{equation}\label{eq:Pthermal}
    \begin{split}
        \rho_\beta =& \bigotimes_\lambda \int \dd[2]{y}_\lambda \frac{1}{\pi\bar{n}_\lambda}e^{-\abs{y_\lambda}^2/\bar{n}_\lambda} \ketbra{y_\lambda},\\
        \bar{n}_\lambda =& \frac{1}{e^{\beta\omega_\lambda}-1}.
    \end{split}
\end{equation}
In order to obtain the finite temperature dynamics we can thus propagate an ensemble of initial states $\ket{\Psi(0)} = \ket{\psi_0}\bigotimes_\lambda\ket{y_\lambda}$ with random, pure coherent bath states $\bigotimes_\lambda\ket{y_\lambda}$. These can be generated from the vaccum with the help of the shift operator
\begin{equation}
    \begin{split}
        D_{\vec{y}} =& \prod_\lambda D_{y_\lambda},\\
        D_{{y_\lambda}} =& e^{y_\lambda a_\lambda^\dagger - y_\lambda^*a_\lambda},\\
        D_{{y_\lambda}}a_\lambda D_{{y_\lambda}}^\dagger =& a_\lambda - y_\lambda.
    \end{split}
\end{equation}
Instead of shifting the initial state, we can (inversely) shift the dynamics and find the following Hamiltonian for the shifted wave vector $\ket{\Psi_{-\vec{y}}} = D_{-\vec{y}}\ket{\Psi}$
\begin{align}
        \partial_t\ket{\Psi_{-\vec{y}}} =& -iD_{-\vec{y}}H_{tot}^ID_{-\vec{y}}^\dagger\ket{\Psi_{-\vec{y}}},\notag\\
            =& -iH^{I,shift}_{tot}\ket{\Psi_{-\vec{y}}},\notag\\
            \label{eq:HtotIshift}
            H^{I,shift}_{tot} \equiv& H_{sys}^{shift} + iB^\dagger(t)L -iB(t)L^\dagger,\\
            H_{sys}^{shift} =& H_{sys} + L\sum_\lambda g_\lambda^* e^{i\omega_\lambda t}y_\lambda^* + L^\dagger\sum_\lambda g_\lambda e^{-i\omega_\lambda t}y_\lambda.\notag
\end{align}
Our transformation has shifted the initial state to the vacuum at the cost of a modified system Hamiltonian. Note, that the $y_\lambda$ appearing in the new system Hamiltonian are drawn from a Gaussian distribution according to Eq.~\eqref{eq:Pthermal}. The new term can thus be interpreted as a stochastic potential depending on the noise $\xi(t) = \sum_\lambda g_\lambda y_\lambda e^{-i\omega_\lambda t}$, which has zero mean and correlation $\mean{\xi(t)\xi^*(s)} = \int_0^\infty \dd{\omega} \bar{n}(\beta\omega)J(\omega)e^{-i\omega(t-s)}$. As Eq.~\eqref{eq:HtotIshift} looks exactly like Eq.~\eqref{eq:HtotI} except for the different system Hamiltonian, we immediately arrive at the nuHOPS equation for finite temperature environments
\begin{equation}\label{eq:shiftedHOPS_finite_T}
    \begin{split}
        \partial_t \ket{\Phi(t)} =& -i \Bigg[H_{sys} + \xi^*(t)L + \xi(t)L^\dagger \\
        & + \omega_c b^\dagger b +i (\mu_{mf}^*(t)L-\mu_{mf}(t)L^\dagger)\\
 & + \sqrt{G} \Big(\left(L - \cavg{L}_0\right)b^\dagger +  \left(L^\dagger - \cavg{L^\dagger}_0\right)b\Big)\Bigg] \ket{\Phi(t)} \\
        &    + \left({z}_t^*L -\kappa b^\dagger b\right) \ket{\Phi(t)}.
    \end{split}
\end{equation}

The reduced system state is then obtained from an average over $\xi(t)$ as well as $z_t$: $\rho_{sys}(t) = {\mathbb{E}}_{\xi(t),z_t}[\braket{0}{\Phi}\!\braket{\Phi}{0}]$.\\
In addition to zero and finite temperature environments, the nuHOPS method paves an easy way to treat a different important kind of initial condition, where only a part of the environment is excited. 
Taking the example from the previous section described by Eq.~\eqref{eq:MasterDephasing}, one could image a situation where the cavity is initially in a coherent state, but still in contact with a zero temperature (Markovian) bath. In the following we give a brief outline how to deal with such initial states.\\
Consider Eq.~\eqref{eq:nonlinear_nmqsd} with an initial state that is an eigenstate of $B(0)$, such that $B(0)\ket{\Psi(0)} = \beta\ket{\Psi(0)}$. 
According to the definition of the auxiliary states \eqref{eq:defAuxStates} we can choose $\mu(0) = \beta$ to obtain the same initial condition for nuHOPS as in the case of a vacuum initial state. 
Now as long as the property $\bra{\vec{z}|}(\partial_tB(t))\ket{\Psi(t)} = -w \bra{\vec{z}|}B(t)\ket{\Psi(t)}$ still holds, we also obtain the same evolution equation for nuHOPS \eqref{eq:shiftedHOPS}. 
In this case we can solve the problem in the same way as the vacuum initial condition just with a different initial condition for $\mu(t)$ and a different noise $\widetilde{z}_t^*$ (arising from the Girsanov transformation). 
A crucial point, however, is that $\bra{\vec{z}|}(\partial_tB(t))\ket{\Psi(t)} = -w \bra{\vec{z}|}B(t)\ket{\Psi(t)}$ does \textit{not} hold for all initial states.
\begin{figure*}
    \centering
    \includegraphics[width=0.315\linewidth]{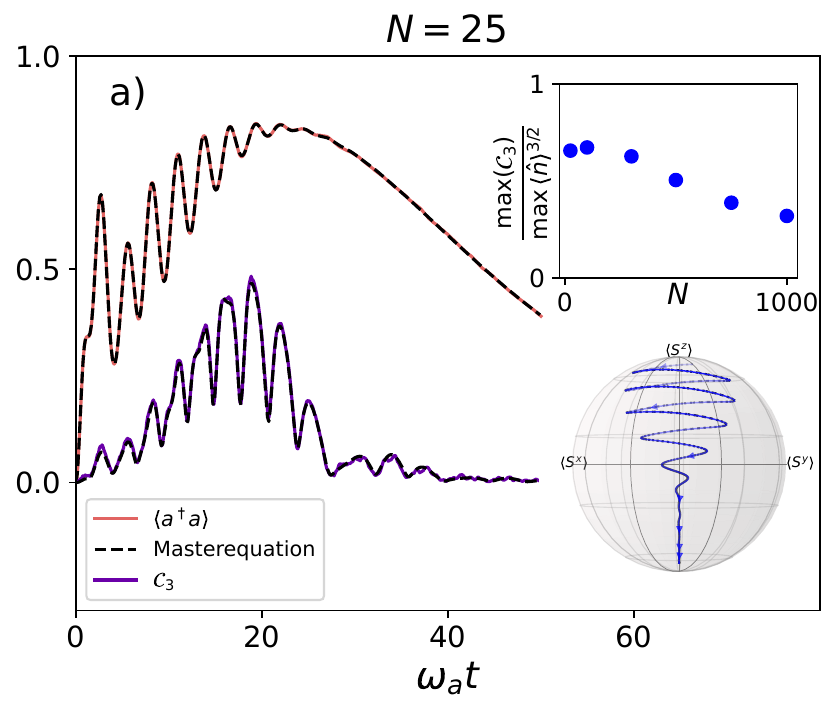}
    \includegraphics[width=0.31\linewidth]{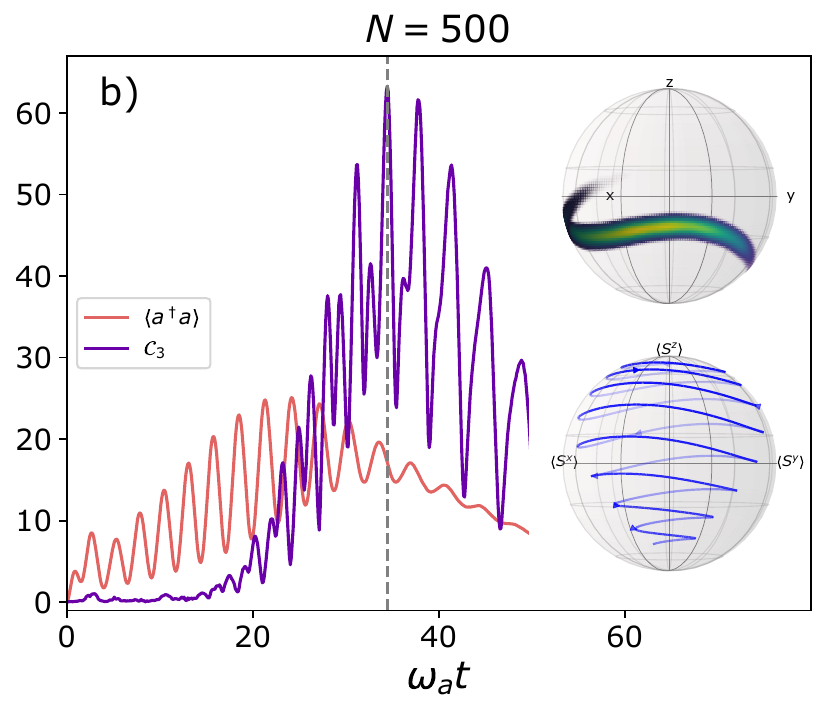}
    \includegraphics[width=0.315\linewidth]{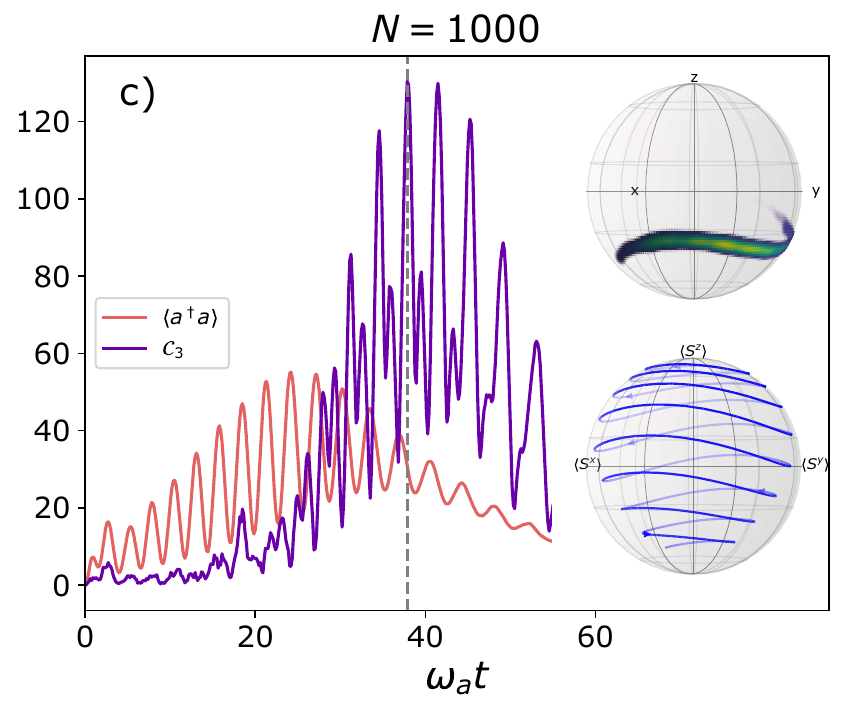}
    \caption{Superradiant decay of an ensemble of atoms for different ensemble sizes. Plots show the number of photons in the cavity as well as the total amount of three-body correlations during the dynamics. Expectation values of the ensemble "spin" are plotted in the lower right inset on a sphere. (a) For $N=25$ a comparison to the exact master equation is possible, which shows perfect agreement. The upper inset shows how the three-body correlations scale with the number of emitters $N$, indicating strongly non-Gaussian states even for large $N$. Plots (b) and (c) show cavity and emitter dynamics for $N=500$ and $N=1000$, respectively. The upper inset shows the spin-Q function of the emitters at the point where $\mathcal{C}_3$ is maximal (indicated by a dashed grey line).}
    \label{fig:Dicke_decay}
\end{figure*}
In Appendix \ref{app:non-thermal} we show that it does hold for an initial coherent state of a cavity $\ket{\psi_{cav}(0)} = \ket{y_0}$ that is coupled to a zero temperature reservoir. 
This initial condition can be taken into account by setting an initial value for the shift $\mu(0) = iy_0/g$ and propagating $\mu(t)$ according to Eq.~\eqref{eq:mu_mf_dgl}. 
By taking an ensemble of different initial shifts also thermal states of the cavity mode could be described. 
We note in passing that these findings do not apply if the cavity is coupled to a finite temperature reservoir.

\section{Superradiant Decay}\label{sec:sr_decay}
Following the previous pedagogical example we now turn to the decay of an ensemble of $N$ emitters inside a high-Q cavity, described by the open Dicke Model \cite{HeppLieb, Dicke_orig_work}
\begin{equation}\label{eq:Master_decay}
    \begin{split}
        \dot{\rho} =& -i[H_D,\rho] + \kappa\left(2a\rho a^\dagger - \{a^\dagger a, \rho\}\right),\\
        H_D =& \omega_a S^z + \omega_c a^\dagger a + \frac{g}{\sqrt{N}}S^x\left(a+a^\dagger\right).
    \end{split}
\end{equation}
The collective decay of an ensemble of emitters has received recent attention in the case of free space \cite{PRL_non_Gauss_driven_cloud,Mink2023Dec,free_space_SR_cumulant_ok} or cavity environments \cite{Stitley_TC}, where special attention is payed to the question whether non-Gaussian correlations can be generated between the emitters or in the output light. 
In this section we demonstrate that the nuHOPS method is ideally suited to tackle such questions in the context of cavity environments. 
This is due to the fact that nuHOPS provides exact solutions which are still numerically feasible in the regime of many emitters, and can also be applied in the good cavity regime $\kappa \approx \omega_a$, going beyond previous works \cite{Stitley_TC}.\\

In Fig.~\ref{fig:Dicke_decay} we show results obtained from a simulation with $\omega_c/\omega_a = 2.5$, $\kappa/\omega_a = 0.5$ and $g/\omega_a = 1.129$ for different emitter numbers $N=25,500,1000$. 
The case of $N=25$ in Fig~\ref{fig:Dicke_decay} a) can also be solved directly with the Master equation \eqref{eq:Master_decay}, where we find perfect agreement. 
The expectation values of the collective "spin" of the emitters are plotted on a sphere in the inset and also clearly agree well with that master equation solution. 
The main plot focuses on the properties of the cavity light. 
To quantify the amount of non-Gaussian correlations in the light we consider the sum of the third order cumulants \cite{Stitley_TC}, which vanish for Gaussian light
\begin{equation}
    \begin{split}
        \mathcal{C}_3 =& \abs{\cavg{aaa}_c} + \abs{\cavg{a^\dagger aa}_c},\\
        \cavg{O_1O_2O_3}_c =& \cavg{O_1O_2O_3} - \cavg{O_1O_2}\cavg{O_3} \\
         &-\cavg{O_1}\cavg{O_2O_3}_c -\cavg{O_1O_3}\cavg{O_2} \\
         &+2\cavg{O_1}\cavg{0_1}\cavg{O_2}\cavg{O_3}.
    \end{split}
\end{equation}
We find clear signs of non-Gaussian correlations in the cavity light and an exact agreement with the direct solution of the Master equation \eqref{eq:Master_decay} for $N=25$. In agreement with previous results for similar systems \cite{Stitley_TC,PRL_non_Gauss_driven_cloud,Mink2023Dec} we find that these non-Gaussian correlations persist also for a large number of emitters as shown in the inset of Fig.~\ref{fig:Dicke_decay} a). The detailed results for $N=500$ and $N=1000$ are shown in Fig.~\ref{fig:Dicke_decay} b), c), where the upper inset shows the spin-Q function of the emitter state at the time where $\mathcal{C}_3(t)$ is maximal. Clearly, significant higher order correlations are present within the emitters as well as in the cavity light.

\section{Conclusions}
We have introduced a nearly unitary version of the hierarchy of pure states, the nuHOPS, a new method for simulating the dynamics of non-Markovian open quantum systems, which remains efficient also for highly excited environments and large system sizes. The
nuHOPS method is based on quantum trajectories, which allows us to reach larger system sizes by propagating state vectors instead of density matrices. Moreover, the freedom to shift the auxiliary oscillator of the hierarchy allows us to minimize the hierarchy depth. 
We have shown how, due to the non-Hermitian terms in the resulting equations, a seemingly well justified Fock-State truncation can lead to rare, sudden instabilities in the propagation and how these can be prevented by using an adaptive basis. The taming of the non-Hermitian  terms in nuHOPS thus comes hand in hand with an improved numerical performance.

Putting our algorithm to the test we have investigated the collective decay of an ensemble of two level emitters in a cavity as described by the (Hepp-Lieb) Dicke model in a regime that was previously not accessible. 
From the numerically exact solution we can conclude that non-Gaussian correlations build up in the cavity as well as within the emitters during the evolution and remain present also for a large number of emitters.\\
In an accompanying letter we apply our method to a generalized (unbalanced) Dicke Model and show how a genuine quantum treatment reveals entanglement in the steady state also at large $N$ and even leads to corrections of the mean-field phase diagram.\\
Going beyond models with a single large spin our method opens the possibility of an exact quantum treatment of more complex systems like individual clumps of atoms in multimode cavities at intermediate sizes. 
Such systems can be used to study the physics of spin glasses in a cavity QED setting \cite{spin_glass_Lev} and have also been considered for associative memory implementations \cite{associativeMemoryPRX}, where the quantum to classical transition is of great interest \cite{Hosseinabadi2023Nov}.
Moreover, for large scale systems, where a simulation will require approximate methods \cite{Hosseinabadi2023Dec}, exact solutions for intermediate sizes can foster understanding by helping to assess the quality of the approximations used.\\
Finally, it remains an interesting question to which extent the improved algorithm can increase the efficiency for problems to which HOPS has been previously applied to, like strongly coupled spin-boson-type models \cite{asymp_entanglement_Richard,valentinBoettcher}, molecular aggregates \cite{Gera2023May} and dissipative many-body systems \cite{HOPS_MPS_Stuart}.
\begin{acknowledgments}
It is a pleasure to thank Valentin Link for enlightening discussions in connection with this work. The authors gratefully acknowledge the computing time made available by the high-performance computer at the NHR Center of TU Dresden. This center is jointly supported by the Federal Ministry of Education and Research and the state governments participating in the NHR (www.nhr-verein.de/unsere-partner).
\end{acknowledgments}
\appendix
\section{Derivation of nuHOPS for a general spectral density}\label{app:derivation}
The following section shows how to set up nuHOPS for an arbitrary bath-correlation function $\alpha(t-s)$. We start the derivation from the non-linear NMQSD equation \eqref{eq:nonlinear_nmqsd} with the replacement \eqref{eq:linear_nmqsd_functional_derivative}:
\begin{equation}\label{eq:NMQSD}
    \begin{split}
        \partial_t \psi_t =& -iH_{sys}\psi_t + \widetilde{z}_t^*L\psi_t \\
        &-(L^\dagger-\cavg{L^\dagger})\int_0^t\dd{s}\alpha (t-s) \frac{\delta}{\delta\widetilde{z}_s^*}\psi_t.
    \end{split}
\end{equation}
First, one approximates the bath correlation function by a sum of exponentials $\alpha(t-s) \approx \sum_j^{k_{max}} G_j e^{-W_j(t-s)}$, where $G_j$, $W_j$ are (complex) fit parameters.
With this sum representation at hand, we now introduce auxiliary states $\psi^{\bvec{e}_j}$ in the spirit of the original HOPS algorithm \cite{HOPS, HOPS_Richard}, analogous to Sec.~\ref{sec:method}. 
For multiple exponentials representing the bath correlation function, a vector index $\bvec{k}$ with dimension $k_{max}$ is required to label all auxiliary states and we denote the unit vectors in direction $j$ by $\bvec{e}_j$. Crucially, for nuHOPS we allow for a shift $\mu_j(t)$ for each auxiliary hiearchy oscillator:
\begin{equation}\label{eq:app_def_aux}
\begin{split}
    \psi_t^{\bvec{e}_j} :=& \frac{-i}{\sqrt{G_j}}\left(\left(\int_0^t\dd{s}G_j e^{-W_j(t-s)}\frac{\delta}{\delta\widetilde{z}_s^*}\right) - \mu_j(t)\right)\psi_t,\\
    \equiv& \frac{-i}{\sqrt{G_j}}\mathcal{D}_j\psi_t,\\
    \psi_t^{\bvec{k}}  =& \prod_{j=1}^{k_{max}}\frac{(-i)^{k}_j}{\sqrt{\bvec{k}_j!G^{\bvec{k}_j}}}\mathcal{D}_j^{\bvec{k}_j}\psi_t.
\end{split}
\end{equation}
Based on this definition we can  express Eq.~\eqref{eq:NMQSD} (neglecting terms which only affect normalization) as
\begin{equation}\label{eq:NMQSD_k}
    \begin{split}
        \partial_t \psi_t^{\bvec{0}}(z^*) =& \left(-iH_{sys} + \widetilde{z}_t^*L - \sum_j^{k_{max}}\mu_j(t)L^\dagger\right)\psi_t^{\bvec{0}}(z^*) \\
        &-(L^\dagger-\cavg{L^\dagger})\sum_j^{k_{max}}\frac{\sqrt{G_j}}{-i}\psi_t^{\bvec{e}_j}.
    \end{split}
\end{equation}
Due to the exponential time dependence under the integral in Eq.~\eqref{eq:app_def_aux} we can make use of the identity $\partial_t\left(\mathcal{D}_j\psi_t\right) = \left(\partial_t \mathcal{D}_j\right)\psi_t + \mathcal{D}_j\dot{\psi}_t = -W_j\mathcal{D}_j\psi_t -(\dot{\mu}_j(t) + W_j\mu_j(t))\psi_t + \mathcal{D}_j\dot{\psi}_t$, which we will show in the following.
Due to causality we must find for $s>t$ that $\frac{\delta}{\delta\widetilde{z}_s^*}\psi_t = 0$, which allows us to extend the upper integral boundary to infinity to find
\begin{equation}
    \begin{split}
        \partial_t&\left(\mathcal{D}_j\psi_t\right)\\
        =& \partial_t\left(\left(\int_0^\infty\dd{s}G_j e^{-W_j(t-s)}\frac{\delta}{\delta\widetilde{z}_s^*}\right) - \mu_j(t)\right)\psi_t,\\
        =& \left(\left(\int_0^\infty\dd{s}(-W_j)G_je^{-W_j(t-s)}\frac{\delta}{\delta\widetilde{z}_s^*}\right) - \dot{\mu}_j(t)\right)\psi_t \\
        &+ \left(\left(\int_0^\infty\dd{s}G_j e^{-W_j(t-s)}\frac{\delta}{\delta\widetilde{z}_s^*}\right) - \mu_j(t)\right)\dot{\psi}_t,\\
        =& -W_j\mathcal{D}_j\psi_t -(\dot{\mu}_j(t) + W_j\mu_j(t))\psi_t + \mathcal{D}_j\dot{\psi}_t.
    \end{split}
\end{equation}
We then define the HOPS state $\ket{\Phi}$ as
\begin{equation}
    \ket{\Phi} = \sum_{\bvec{k}}\psi_t^{\bvec{e}_j}\bigotimes_{j=1}^{k_{max}}\ket{{k}_j}.
\end{equation}
With the help of the previously shown identity we find
\begin{equation}\label{eq:nuHOPSgeneral}
    \begin{split}
        \partial_t \ket{\Phi} =& \left(-i H_{sys} + \widetilde{z}_t^*L\right)\ket{\Phi} \\
        &-\sum_{j=1}^{k_{max}}\Bigg(\mu_j(t)L^\dagger + W_jb_j^\dagger b_j\\
        &+\quad i\sqrt{G_j} \left(L - \frac{\dot{\mu_j} + W_j\mu_j}{G_j}\right) b_j^\dagger\\
        &+\quad i\sqrt{G_j}\left(L^\dagger - \cavg{L^\dagger}_0\right)b_j\Bigg)\ket{\Phi},\\
        \equiv& \mathcal{G}(z^*)\ket{\Phi}
    \end{split}
\end{equation}
where $b_j$($b_j^\dagger$) are the annihilation (creation) operators for auxiliary mode $j$ and we use the notation $\cavg{L^\dagger}_0 = \langle\Phi\ket{0}L^\dagger\bra{0}\ket{\Phi}/\langle\Phi\ketbra{0}\Phi\rangle$ to emphasize that the expectation value is taken with respect to the zeroth order state only.
Since we consider zero temperature, all auxiliary states are initially in the vacuum state.
The initial state thus reads $\ket{\Phi(0)} = \ket{\psi_0}\otimes\ket{0}$.
As mentioned before, the reduced density matrix can be obtained from the solution of this equation by taking the average $\rho_{sys} =\mean{\bra{\bvec{0}}\ketbra{\Phi}{\Phi}\ket{\bvec{0}}}$, where $\ket{\bvec{0}}$ represents the ground state of the auxiliary oscillators.

\section{Derivation of the optimal shift}\label{app:optimal_shift}
In the following we turn to the derivation of the optimal shift. 
In Eq.~\eqref{eq:shiftedHOPS},\eqref{eq:nuHOPSgeneral} the functions $\mu_j(t)$ are still arbitrary. 
We now proceed to exploit this freedom in order to decrease the number of hierarchy states (Fock states of the auxiliary oscillators) required to accurately simulate Eq.~\eqref{eq:nuHOPSgeneral}.
A Fock space expansion of the auxiliary oscillators is most efficient if their quantum state is centered near their ground states, that is $\cavg{b_j}=0$ for all $j$. 
The time-evolution of $\cavg{b_j}$ is determined by Eq.~\eqref{eq:nuHOPSgeneral} and with the use of the abbreviation ${m}_j := \frac{\dot{\mu_j} + W_j\mu_j}{G_j}$ we obtain
\begin{align*}
\partial_{t}\langle b_k\rangle=&\partial_{t} \frac{\langle\Phi|b_k| \Phi\rangle}{\bra{\Phi}\ket{\Phi}}, \\
=&\frac{\langle\Phi|b_k \mathcal{G}| \Phi\rangle+\langle\Phi|\mathcal{G}^{\dagger}b_k \mid \Phi\rangle}{\bra{\Phi}\ket{\Phi}}-\frac{\langle\Phi|{b_k}| \Phi\rangle \partial_{t}\bra{\Phi}\ket{\Phi}}{\bra{\Phi}\ket{\Phi}^2} \\
=& \cavg{b_k\mathcal{G} + \mathcal{G}^\dagger b_k} - \cavg{b_k}\cavg{\mathcal{G}+\mathcal{G}^\dagger}
\end{align*}
For the expectation values we find
\begin{align*}\label{eq:app_exp_values_deriv}
\cavg{b_k\mathcal{G} + \mathcal{G}^\dagger b_k}=&\widetilde{z}_t^*\cavg{b_k L} + \widetilde{z}_t\cavg{b_k L^\dagger} - W_k\cavg{b_k}\\
&+i\sqrt{G_k}(\cavg{L}_0-m_k) \\
&-\sum_j^{k_{max}}\Bigg(\mu_j\cavg{b_k L^\dagger} - \mu_j^*\cavg{b_k L} \\
&\quad+2\operatorname{Re}(W_j)\cavg{b_j^\dagger b_j}\\
&\quad+ i\sqrt{G_j}(m_j^* -\cavg{L^\dagger}_0)\cavg{b_k b_j} \\
&\quad-i\sqrt{G_j}(m_j -\cavg{L}_0)\cavg{b_j^\dagger b_k}\Bigg),\\
\cavg{\mathcal{G} + \mathcal{G}^\dagger} =& \widetilde{z}_t^*\cavg{L} + \widetilde{z}_t\cavg{L^\dagger} \\
&-\sum_j^{k_{max}} \Bigg(\mu_j \cavg{L^\dagger} + \mu_j^*\cavg{L} + 2\operatorname{Re}(W_j)\cavg{b_j^\dagger b_j} \\
&\quad+i\sqrt{G_j}(m_j^* - \cavg{L^\dagger}_0)\cavg{b_j} \\
&\quad- i\sqrt{G_j}(m_j - \cavg{L}_0)\cavg{b_j^\dagger}\Bigg)
\end{align*}
As each $m_j$ contains the derivative of $\mu_j$ the above equations can be rearranged to find coupled differential equations for the optimal $\mu_j(t)$. However, we found that, especially for the Dicke model, a much simpler differential equation works just as well. Using a mean-field approximation to split the expectation values above we arrive at
\begin{equation*}
    \partial_t \cavg{b_k} \approx -W_k\cavg{b_k} +i\sqrt{G_j}\cavg{L}_0 -\frac{i}{\sqrt{G_k}}\left(\dot{\mu}_k + W_k\mu_k\right).
\end{equation*}
To ensure that $\cavg{b_k}\approx 0$ is a stable fixed point of the dynamics we demand $\partial_t\cavg{b_k} = -W_k\cavg{b_k}$ (note that by construction $\cavg{b_k}(t=0) = 0$), which is satisfied if
\begin{equation}\label{eq:mfShiftAppendix}
    \dot{\mu}_k = -W_k\mu_k + G_k\cavg{L}_0
\end{equation}
These mean-field-optimal $\mu_k$ correspond exactly to the stochastic process shifts \eqref{eq:def_stocproc}, which are already calculated during the HOPS evolution to ensure efficient sampling. Eq.~\eqref{eq:nuHOPSgeneral} can thus be used without additional overhead, while still resulting in a near optimal representation of the auxiliary oscillators, and moreover, in a near-unitary evolution of the HOPS state.

\section{Application to cavity-QED systems}\label{app:HOPSforcavity}
We briefly sketch why a Lindblad Master equation describing a damped cavity as in Eq.~\eqref{eq:MasterDephasing} and \eqref{eq:Master_decay} corresponds to a system described by Eq.~\eqref{eq:Htot} with an exponentially decaying correlation function. This is a well-known relation and the basis for pseudo-mode methods in open system dynamics.\\

The Lindblad Master equation is obtained from eliminating a Markovian bath, which is coupled to the cavity mode. We thus start from the total Hamiltonian containing the system, the cavity mode at frequency $\omega_c$ and the Markovian bath explicitly.
\begin{equation}
\begin{split}
    H_{tot} =& H_{sys} + g(La^\dagger + L^\dagger a) + \omega_c a^\dagger a \\
    &+ \sum_j \left(\Omega_jc_j^\dagger c_j + d_j(ac_j^\dagger + a^\dagger c_j)\right),\\
    \equiv&   H_{sys} + g(La^\dagger + L^\dagger a) + H_{CM}.
\end{split}
\end{equation}
Here $c_j$ ($c_j^\dagger$) are the annihilation operators describing the continuum of modes in the bath. As the bath is Markovian its correlation function is taken as a delta-function 
$\sum_j d_j^2 e^{-i\Omega_j(t-s)} = 2\kappa \delta(t-s)$. We now switch to an interaction picture with respect to the cavity and bath Hamiltonian $H_{CM}$, and find
\begin{equation}\label{eq:app_evolution_int_pic}
\begin{split}
    \partial_t\ket{\Psi^I(t)} =& -i\left(H_{sys} + g(L^\dagger a^I(t) + L (a^I)^\dagger (t))\right)\ket{\Psi^I(t)}.
\end{split}
\end{equation}
The cavity annihilation operator in the interaction picture satisfies the equation of an operator Ornstein-Uhlenbeck process,
\begin{align} \label{eq:app_ai_dot}
   \dv{a^I}{t}  
    =& -(i\omega_c + \kappa)a^I(t) -i \zeta(t),
\end{align}
where $\zeta(t) = \sum_j d_j e^{-i\Omega_j t} c^I(0)$ is operator white noise with $[\zeta(t),\zeta^\dagger(s)] = 2\kappa\delta(t-s)$.
We obtain the solution
\begin{equation}
    \label{eq:app_a_int}
    a^I(t) = e^{-(i\omega_c+\kappa)t}a - i\int_0^t\dd{s}e^{-(i\omega_c +\kappa)(t-s)}\zeta(s).
\end{equation}
Comparing Eq.~\eqref{eq:app_evolution_int_pic} to \eqref{eq:HtotI} we can identify $B(t)$ with $ga^I(t)$. The corresponding correlation function at zero temperature is thus the exponentially decaying correlation function of the Ornstein-Uhlenbeck process,
\begin{align}
    \cavg{B(t)B^\dagger(s)}
        =&g^2  \cavg{a^I(t)a^{I\dagger}(s)},\notag\\
        =& g^2e^{-i\omega_c(t-s)-\kappa\abs{t-s}}.
\end{align}
Thus, a damped cavity mode can be simulated with nuHOPS by using an exponentially decaying correlation function.

\section{Non-thermal initial states}\label{app:non-thermal}
In Sec.~\ref{sec:finite_T} we have shown how nuHOPS can be used to efficiently deal with any initial state that is an eigenstate of the operator $B(0)$, as long as $\bra{\vec{z}|}(\partial_tB(t))\ket{\Psi(t)} = -w \bra{\vec{z}|}B(t)\ket{\Psi(t)}$ still holds. 
In the following we show that this is the case in a damped cavity, when the cavity is in a coherent state $\beta$, and the Markovian bath is still at zero temperature.
The initial state thus reads $\ket{\Psi(0)} = \ket{\psi_0}\otimes\ket{\beta}\otimes\ket{0}$. In Appendix \ref{app:HOPSforcavity} we have shown that for a damped cavity the operator $B(t)$ corresponds to $ga^I(t)$, where $a^I(t)$ is the cavity annihilation operator. With Eq.~\eqref{eq:app_ai_dot} we find
\begin{align}\label{eq:app_func_deriv_wn}
    \bra{\vec{z}|}(\partial_tB(t))\ket{\Psi(t)} =& g\bra{\vec{z}|}(\partial_ta^I(t)))\ket{\Psi(t)}\notag,\\
    =& g \bra{\vec{z}|}-(i\omega_c + \kappa)a^I(t) -i\zeta(t)\ket{\Psi(t)},
\end{align}

The last term can be evaluated easily: $|\Psi(t)\rangle$ depends on $\zeta(s)$ with arguments $s<t$ only. Thus, as $\zeta(t)$ is white noise, time evolution up to time t and $\zeta(t)$ commute, and we can apply the latter to the initial state $|\Psi(0)\rangle$, which gives zero.
Thus the last term in Eq.~\eqref{eq:app_func_deriv_wn} vanishes and indeed, we find 
\begin{align}
 \bra{\vec{z}|}(\partial_tB(t))\ket{\Psi(t)} = -(i\omega_c + \kappa)\bra{\vec{z}|}B(t)\ket{\Psi(t)}.  
\end{align}

\bibliography{literature}

\end{document}